\documentclass[twcolumn]{aastex631}

\usepackage{lineno}
\usepackage{graphicx}

\shorttitle{Jet/shear layer }
\shortauthors{Raga et al.}
\graphicspath{{./}{figures/}}

\begin{document}

\title{HH~270/110 as a jet/shear layer interaction}


\author[0000-0002-0835-1126]{A. C. Raga}
\affiliation{Instituto de Ciencias Nucleares \\
  Universidad Nacional Aut\'onoma de M\'exico \\
  Ap. 70-543, Ciudad Universitaria, CDMX 04510, M\'exico}

\author[0000-0002-6296-8960]{A. Noriega-Crespo$^*$}
\affiliation{Space Telescope Science Institute \\
3700 San Martin Dr.\\
Baltimore, MD 21211, USA}

\author[0000-0003-3974-8449]{A. Castellanos-Ram\'\i rez}
\author[0000-0003-3863-7114]{J. Cant\'o}
\affiliation{Instituto de Astronom\'\i a \\
  Universidad Nacional Aut\'onoma de M\'exico \\
  Ap. 70-658, Ciudad Universitaria, CDMX 04510, M\'exico}
  
\author[0000-0001-5653-7817]{H. Arce}
\affiliation{Dep. of Astronomy, Yale University\\
  New Haven, CT 06511m USA}

\author[0000-0001-7943-9961]{M. Lebr\'on}
\author[0000-0003-0759-4991]{J.L Morales Ortiz}
\affiliation{Univ. of Puerto Rico-R\'\i o Piedras \\
  General Studies Faculty, Dep. of Physical Sciences \\
  San Juan, Puerto Rico, USA}

\affiliation{Univ. of Puerto Rico-R\'\i o Piedras \\
  Natural Sciences Faculty, Dep. of Physics \\
  San Juan, Puerto Rico, USA}
\author[0009-0007-1261-4893]{A. N. Ortiz Capeles}
\author[0000-0003-0759-4991]{C. A. Pantoja}
\affiliation{Univ. of Puerto Rico-R\'\i o Piedras \\
  Natural Sciences Faculty, Dep. of Physics \\
  San Juan, Puerto Rico, USA}

\begin{abstract}
  New observations obtained with JWST of the proto-stellar HH~270 jet and
  the ``deflected'' HH 110 system, show that
  HH~110 has a morphology of a series of distorted
  working surfaces. These working surfaces appear to be ``deflected versions''
  of the heads of the incident, HH~270 jet. We compute a series of
  3D numerical simulations, in which we explore the possible parameters
  of a shearing environment that could give origin to the deflection
  of HH~270 into the HH~110 flow. We find that we need a quite high
  sideways velocity for the streaming environment (of $\sim 30$~km~s$^{-1}$)
  in order to produce the complex structure observed in HH~110. This
  high velocity would be possible in an environment which has been
  strongly perturbed by the passage of other outflows.
\end{abstract}


\section{Introduction}

HH~110 (discovered by Reipurth \& Olberg 1991) is a Herbig-Haro (HH)
jet with no identified source along the outflow axis. It is also
a somewhat unusual outflow, with a not so well collimated beam of
turbulent appearance. Reipurth et al. (1996) proposed that a newly
discovered jet (HH~270) might be deflected into the direction of
HH~110 through a collision with a dense, environmental structure.
The higher HH~270 and lower HH~110 proper motions (studied later
in more detail by L\'opez et al. 2005 and Kajdic et al. 2012)
are consistent with this ``deflected jet'' scenario.

Since the discovery of HH~110, many more ``deflected jet'' systems
have been observed. For example, some star formation regions with
many outflows (e.g., Raga et al. 2013; Nony et al. 2020) show
several deflected jets, which are possibly the result of collisions
with dense environmental condensations or with other outflows. Therefore,
there are now many candidates for exploring the associated processes.

A comparison of the radial velocities (of $\sim 25\to 50$~km~s$^{-1}$,
see Riera et al. 2003a, b) and the proper motions (of $\sim 100$~km~s$^{-1}$,
see Kajdic et al. 2012) indicates that HH~110 lies close to the
plane of the sky. Interestingly, though HH~110 was fully covered
by the observations of Riera et al. (2003b), their Fabry-P\'erot map
did not have the sensitivity to detect HH~270, and as far as we
know there are no published radial velocities of this jet. However,
the high proper motions of HH~270 ($\sim 250$~km~s$^{-1}$, see Kajdic
et al. 2012) and the fact that optical images of
the source region show an  a 'disk shadow' dark
nebula (Figure 8 of Kajdic et al. 2012) indicates
that the HH~270 axis is also close to the plane of the sky. Therefore,
we are looking at a jet deflection taking place close to the plane
of the sky, which to some extent simplifies its interpretation.

Noriega-Crespo et al. (1996) showed that the H$_2$~2.124$\mu$m emission
of HH~110 is restricted to a narrow layer located to the W of the optical
(atomic/ionic) emission. They interpreted this interesting feature
in terms of an analytic, turbulent mixing layer model, with the mixing
layer corresponding to the sideways interaction of the HH~110 jet with a
molecular environment. This narrow H$_2$ emission is better seen
in the Spitzer space telescope image presented by Kajdic
et al. (2012, their Figure 10), and in a new JWST image which
we show in the present paper (see \S 2). As could have been
expected, this JWST image shows enlightening new details of the
morphology of the H$_2$ emission.

The interpretation of the IR and optical emission of HH~110 in
terms of a mixing layer model (Noriega-Crespo et al. 1996) is
not consistent with other models that have been developed for
the HH~270/110 flow. The models for producing the deflection have
been centered on a scenario in which HH~270 is deflected in
a collision with a dense cloud, and then propagates within
a surrounding, low density environment (de Gouveia Dal Pino 1999;
Raga et al. 1995; Raga et al. 2002).

These jet/cloud collision models can produce the deflection
observed in HH~270/110, and also reproduce the decrease in
proper motions observed when going from the incident (HH~270)
to the deflected (HH~110) parts of the outflow. However, the
success at reproducing the observations is only partial:
\begin{itemize}
\item the deflected jet structure survives only for a relatively
  short time before the incident jet starts directly boring
  a hole into the dense obstacle (Raga et al. 1995),
\item even if the incident jet is variable (with successive
  working surfaces), the deflected outflow does not show
  the complex spatial structure of the HH~110 jet (see Raga et al.
  2002),
\item the mixing layer between the jet and the molecular cloud
  is limited to a region close to the point of collision between
  the jet and the cloud, and does not extend along the deflected
  jet (as assumed by Noriega-Crespo et al. 1996 for explaining the
  H$_2$ emission of HH~110, see above).
\end{itemize}

Raga et al. (2023) considered an alternative model for producing
a jet deflection, in which a jet meets a lateral velocity shear
layer present in the environment. This flow produces a deflected
jet which is in lateral contact with the streaming environment, and
naturally has a spatially extended mixing layer between the shocked jet and
(streaming) environmental material. Also, the deflected jet beam
exists for the whole duration of the outflow phenomenon. An interpretation
of HH~110 as a deflection on collision with a second outflow was
first suggested by Kajdic et al. (2012), and is of course quite
similar to the jet/shear layer interaction scenario of Raga et al. (2023).

Another alternative is that HH~110 results from a pulsed jet (HH~270) that is precessing. The pulses hit a stationary medium (dense cloud), and given the appropriate timing for the ejection pulses, the precession period, and the ejection velocity, it seems possible to account for the observed geometry of HH~110. However, this “precessing machine-gun” model predicts a velocity field that is opposite to what is observed (see e.g. Raga and Biro 1993; Raga, Canto and Biro 1993). The ejection flow has no angular momentum, so the interaction between the ejection bullets and a stationary medium will produce clumps that move radially away from the source. However, the clumps that constitute HH~110 seem to move along the object and not in a radial direction away from the source (see Figure 5).

In the present paper we
extend the model of Raga et al. (2023) to the case of a variable
ejection velocity jet, and concentrate on parameters appropriate
for the HH~270/110 system. The paper is organized as follows.
Section 2 presents a comparison between an optical (H$\alpha$+[S~II])
and a new F460M JWST image of the HH~270/110 system, discussing the
implications of the new IR image. Section 3 discusses the
choice of appropriate parameters for modeling HH~270/110
in terms of the interaction of a variable jet with an environmental
velocity shear. Section 4 describes the setup of the 3D numerical
simulations, and the parameters of the 8 models that are included
in this paper. Section 5 presents the results of the numerical
simulations, and compares them with the observed characteristics
of HH~270/110. Finally, Section 6 discusses the implications
of our results, and the choices of parameters that give results
that better agree with the observed structure of HH~270/110.
\vskip 0.5in

\section{Observations of HH~270/110}

In this section, we present a new image of HH~270/110 obtained with
the James Webb Space Telescope (JWST, Gardner et al. 2023; Rigby et
al. 2023), obtained as part of the  Guaranteed Observing Time
(GTO) 1293 program (Noriega-Crespo et al. 2017). The observations
were obtained with both NIRCam (Rieke et al. 2023) and MIRI
(Wright et al. 2023) on October 2022 in parallel mode, demonstrating the versatility of the JWST capabilities of observing with more than one instrument at the time. We present a 3$\times$2 (15\%\ overlap) section of the mosaic of the HH~270/HH~110 region obtained with F460M filter using a SHALLOW4 readout pattern, with
4 groups and 3 dithers and a total exposure time of 612 seconds per tile. The final image covers
$\sim 0.7\arcmin \times 0.8\arcmin$ with an angular resolution of 
approximately 126 milli-arcsec.  The F460M image corresponds to the final calibrated product of the standard STScI JWST pipeline (Level 3) and no further processing has been applied, that for the purpose of this work, that is mostly focus on the numerical simulations stimulated by these recent JWST observation, has been enough.
A detailed description of the observations will be presented elsewhere (Noriega-Crespo et al. in preparation)

The NIRCam F460M filter bandpass encompasses the 0-0 S(9) H2 line at
4.695 $\mu$m, as well as several of the CO (1-0) R-branch components (see.e.g. Ray et al. 2023). The S(9) emission line was successfully used by the Spitzer Space Telescope and the IRAC instrument in its Channel 2 (4.5 $\mu$m) to detect 
shock excited emission arising from proto-stellar outflows.
Figure 1 shows a  comparison of the IRAC Channel 2 (left)
and the NIRCam F460M over a 3$\times$ 3 arcminute FOV (right). The NIRCam observations certainly suggest that the working surfaces of the outflow, the bowshocks, are following a curve trajectory with a sharper turn on what is known as HH~110.
Furthermore the last structure at the bottom of the image does seem even more displaced than the rest of the of the flow, and looking at the IRAC image, less aligned with the HH~451 outflow axis, indicating that something else (perhaps precession) is also taking place.

Figure 2 shows two images of the HH~270/110 system:
\begin{itemize}
\item the red [S~II]+H$\alpha$ image of Kajdic et al. (2012),
  obtained with the Subaru telescope in Hawaii,
\item a new JWST image obtained with the F460M filter (with
  a $\lambda_0= 4.624$~$\mu$m central wavelength
  and a $\Delta \lambda=0.228$~$\mu$m bandwidth), in which
  the detected emission is probably dominated by H$_2$ lines.
\end{itemize}
The two images have been rotated 122$^\circ$ (counterclockwise), so that
the HH~270 axis is approximately vertical, and the displayed
region has a horizontal size of $300''$ (corresponding
to $1.8\times 10^{18}$~cm=0.58~pc at a distance of 400~pc to
HH~110).

The region around the outflow source is seen as a dark band
in the optical image, and as an extended reflection nebula
in the F460M map (in the lower, central region of the frames).
The knots along the (vertical) HH~270 flow are seen as two broken
up bow-like structures in the two frames. HH~110 is seen as
a broadening, structured emission region in the optical image
and as a much narrower, broken up flow in the F460M image. Finally,
in the left region of the optical map there is a dark, high
extinction region, from which HH~270 appears to emerge. This
dark region is not seen in the F460M map (see Figure 1).

\begin{figure}
\epsscale{1.1}
\plotone{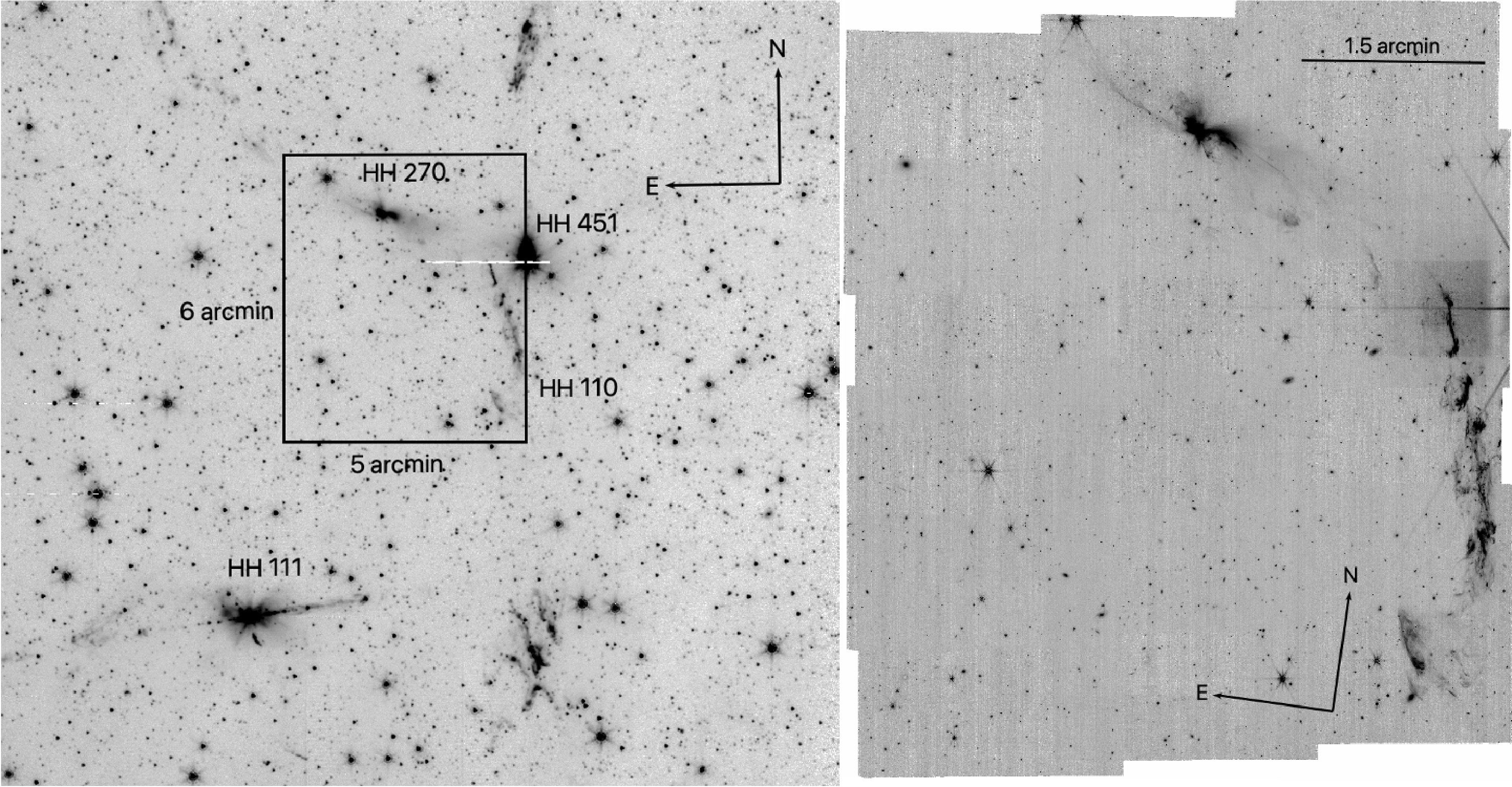}
\caption{Spitzer IRAC Channel 2 map at 4.5$\mu$m (left). The box corresponds to the region covered by the JWST NIRCam
  F460M image (right), at a comparable wavelength. The NIRCam FOV covers 6$\times$ 5 arcminutes, and contains the HH~270/110 flow that clearly shows the HH~270 bowshocks merging towards HH~110 (see also Fig. 10).
\label{fig1}}
\end{figure}

\begin{figure}
\epsscale{0.7}
\plotone{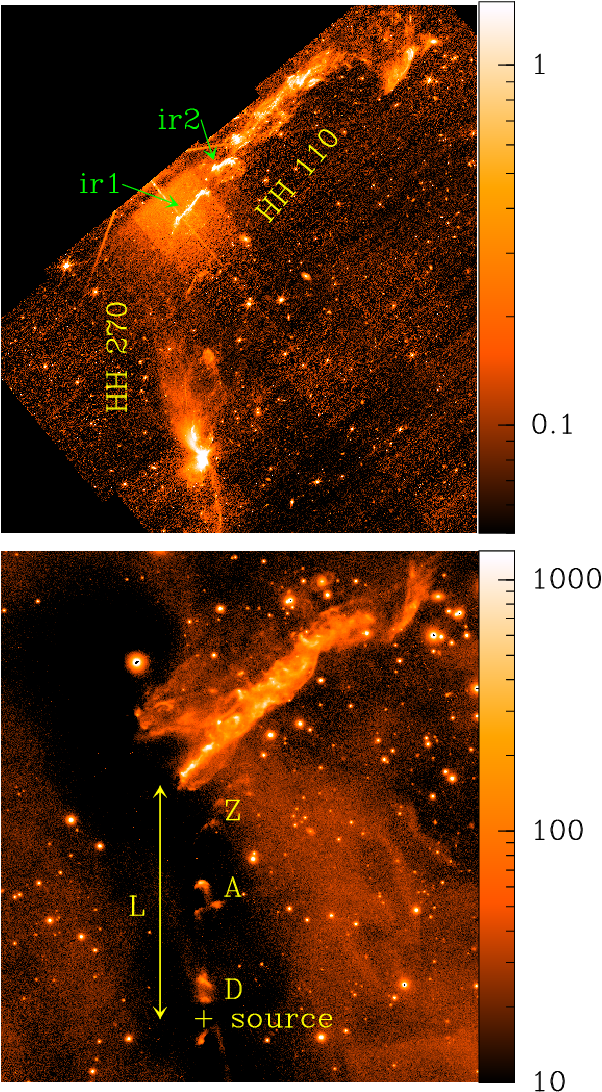}
\caption{JWST F460 (top) and a Subaru [S~II] + H$\alpha$ image (bottom)
  of the HH~270 and HH~110 jets. The images have been rotated (a
  122$^\circ$ counterclockwise rotation from a North up orientation) so
  that the HH~270 jet axis is approximately vertical. The images
  are shown with the linear colour scales on the right, and have
  a horizontal size of $300''$ ($1.8\times 10^{18}$~cm=0.58~pc at
  a 400~pc distance). In the top frame, the regions of the
  HH~270 and HH~110 flows are labeled, as well as the
  two ``distorted bows'' (ir1 and ir2, with green labels) discussed
  in section 2. The approximate position of the source of the
  outflow system, as well as the length $L$ and three knots (D, A and Z)
  of HH~270 are identified in the bottom frame (see Figs 9 and 10).
\label{fig2}}
\end{figure}

The high angular resolution of the F460M image shows that HH~110
has a structure of successive ``distorted arc'' structures. These
structures have sizes and separations which are similar
to the bows in the HH~270 flow. This morphology could be consistent
with a situation in which the HH~270 working surfaces are advected
sideways by an environment with a strong velocity shear
(starting in the region in which HH~110 begins). This remarkable
morphology was not evident in the previous, lower resolution
IR images of HH~110 (see, e.g., Noriega-Crespo et al. 1996).

\section{The Multiple working surfaces from HH~270}

From these images we see that the HH~270 jet has a length $L\approx 150''$
(measured from the outflow source to its intersection with HH~110,
see Figure 2), corresponding to $\approx 9.2\times 10^{17}$~cm
(at 400~pc). The separation between the HH~270 heads
(knots A and Z of Kajdic et al. (2012), see Figure 2)
is $\Delta x\approx 50''$ ($3.1\times 10^{17}$~cm).

If we assume that these heads are the result of an ejection velocity
time-variability of the source, we can combine the separation
$\Delta x$ between working surfaces with their proper motion velocity
$v_j\approx 250$~km~s$^{-1}$ (see Kajdic et al. (2012)), to obtain
a variability period $\tau\approx \Delta x/v_j\approx 400$~yr.

We now furthermore assume that we have a sinusoidal ejection
velocity variability of the form:
\begin{equation}
  v(t)=v_j+\Delta v\,\sin\left(\frac{2\pi t}{\tau}\right)\,,
  \label{v}
\end{equation}
where $v_j$ is the mean ejection velocity, $\Delta v$ is the
velocity half-amplitude, $\tau$ is the period and $t$ the time
of the ejection.

The ejection velocity variability leads to the formation
of ``internal working surfaces'' that travel down the jet beam.
For a smooth ejection variability, these working surfaces form at
a finite distance $x_c$ from the source (as described by Raga et al.
1990 and several later papers). This distance of working surface
formation can be used to estimate $\Delta v$.

If the first
working surface is knot A of HH~270 (see Kajdic et al. (2012)),
we obtain an upper limit $x_0\approx 70''$ ($4\times 10^{17}$~cm
at 400~pc) for $x_c$. Combining this value with the proper motion
velocity of the knots (see above) we obtain
a lower limit for the half-amplitude $\Delta v$ of the ejection
time-variability:
\begin{equation}
  \Delta v_{min}=\frac{\tau v_j^2}{2\pi x_c}\approx 30\,{\rm km\,s^{-1}}\,.
  \label{vmin}
\end{equation}
This is equation (7) of Raga \& Cant\'o (1998).

If the first working surface were knot D of Kajdic et al. (2012), we
would obtain a lower estimate of $\approx 100$~km~s$^{-1}$ for $\Delta v$.
However this optical knot appears to be part of the reflection
nebula around the outflow source in the F460M image. In the following,
we adopt a $\Delta v=50$~km~s$^{-1}$ half-amplitude for the variability
of the HH~270 ejection.

In the optical and IR images one can also measure the angle between
the HH~270 and HH~110 axes. This deflection angle is of $\delta \approx
41^\circ$.

The estimate of the velocity variability
amplitude, of the jet velocity, the deflection angle,
and the measurements of the lengths of HH~270 and HH~110 are
of course dependent on projection effects. However, as argued in
\S 1, the HH~270 and HH~110 flows appear to lie close to the
plane of the sky, so that these projection effects are probably not
important. The jet parameters are further discussed in the following
section.

\section{The variable jet/shear layer interaction problem}

We consider a variable jet flow that punches through a plane environmental
velocity shear layer. Unfortunately, this problem has a large number of
free parameters associated with:
\begin{enumerate}
\item the properties of the time-dependent ejection,
\item the characteristics of the environment around the
  outflow source,
\item the structure of the environmental shear layer
and the properties of the post-shear layer environment,
\item the direction of the jet relative to the surface of
  the shear layer.
\end{enumerate}
In the present paper, we carry out a limited exploration
of this parameter space, which we describe in the following subsections.

We first fix the properties of the time-dependent ejection so as to
produce internal working surfaces with velocities and spacings similar to
the ones of the HH~270 incident jet (see section 2). To this
effect, we choose a sinusoidal ejection velocity variability
with a mean velocity $v_j=250$~km~s$^{-1}$, a half-amplitude
$\Delta v=50$~km~s$^{-1}$ and a $\tau=400$~yr
period. We choose a time-independent $n_j=100$~cm$^{-3}$ ejection density,
and a $T_j=2000$~K temperature.

The jet is initially cylindrical, with a $r_j=5\times 10^{16}$~cm
radius, which gives a configuration that produces internal working
surfaces of widths similar to the bow shocks along the HH~270 jet (see
section 2). Taking the mean jet velocity and the initial density
(see above), with this chosen radius we obtain an average mass loss
rate of $\approx 2.5\times 10^{-7}$~M$_\odot$yr$^{-1}$ for the ejection.

For the stationary environment into which the jet is initially traveling,
we choose a density and a temperature such that this region is
in pressure balance with the flowing environment on the other side
of the environmental velocity shear layer. This pressure condition
is necessary for maintaining a relatively stable shear layer. We find
that strong perturbations rapidly appear in shear layers between
two regions with unequal pressure.

We assume that the initial environment is composed of two regions,
one at rest with respect to the outflow source (see \S\S 3.2) and
the second one with a velocity $v_a$, parallel to the plane boundary
between the two regions. On the upstream boundary, an inflow condition
with the variables of the flowing region is imposed at all times.

The transition between the two regions has a width equal to the
computational grid resolution in the initial condition, and on the
inflow boundary. As the flow evolves, the interaction region
develops a width of a few grid points in the downstream region
of the environmental flow.
For the post-shear layer region, we explore a few values
for the velocity $v_a$ and the density $n_a$ of the flowing
environment. In all of our models, we have assumed that
this region has a $T_a=100$~K temperature.

We have restricted our models to the case in which the jet
velocity $v_j$ and the environmental velocity $v_a$ lie on
a plane perpendicular to the plane of the environmental
shear layer. For this configuration, we compute flows with
different angles $\phi$ between the $v_j$ and $v_a$ directions.

The general geometry of the flow is shown in the schematic diagram
of Figure 3. The $xz$-plane contains $v_j$ and $v_a$, with the
jet initially traveling along the $z$-axis. The interaction
of the jet with the streaming environment (present in the upper
part of the region shown in Figure 3) produces a compression
and a deviation from the initial jet beam into a deflected
jet/streaming environment interaction region (shown with the
thick, red curve). The jet is initially deflected an
angle $\delta$ from its original direction, and then continues
on a curved trajectory as more momentum from the streaming environment
is incorporated into the deflected flow.

Raga et al. (2023) studied the case of a jet with constant
ejection velocity and density, and derived the relation
\begin{equation}
  \tan \delta=\frac{\sin\phi}{1/p+\cos\phi}\,,
  \label{del}
\end{equation}
with
\begin{equation}
  p=\sqrt{\frac{\rho_a v_a^2}{\rho_j v_j^2}}\,,
  \label{p}
\end{equation}
being the square root of the streaming environment to jet ram-pressure
ratio and $\phi$ the angle between the jet axis and the environmental
shear surface (see Figure 3). From Figure 3 of Raga et al. (2023)
it is clear that in order
to obtain $\delta\sim 45^\circ$ (as necessary for modeling the
HH~270/110 flow, see \S 2), it is necessary to
have $p\sim 1$, as well as  $\phi\sim 90^\circ$ (i.e., a
high ram-pressure streaming environment and flowing approximately
perpendicular to the initial jet direction).

The models of the present paper have a variable ejection velocity
jet, so that the results of Raga et al. (2023) do not apply in
a direct way. However, we find that the deflection of the beam of
a variable jet (which has generated internal working surfaces
before hitting the streaming environment) still follows
equation (\ref{del}) in an approximate way. We therefore compute
models that straddle the $p=1$ and $\phi=90^\circ$ values, in
order to obtain substantial deflections.

\begin{figure}
\epsscale{0.5}
\plotone{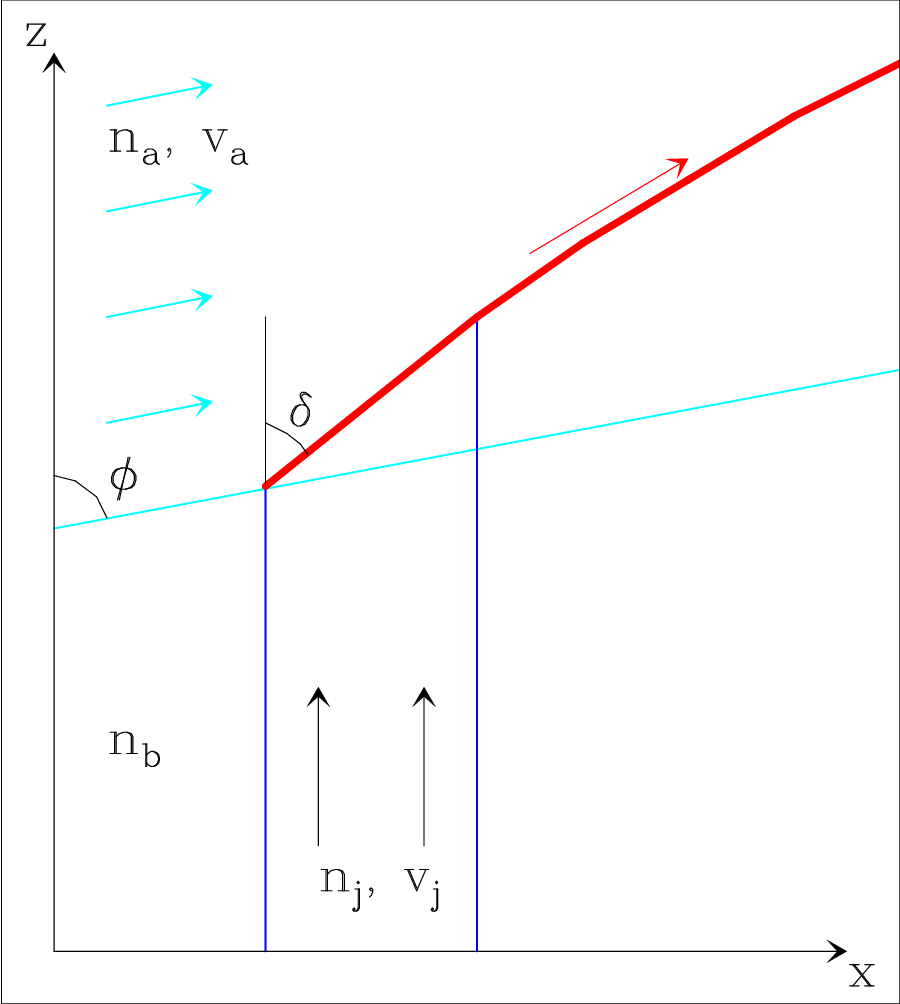}
\caption{Schematic diagram showing the interaction of a jet
  that crosses the boundary between a stationary and a streaming
  region of the surrounding environment. The initial jet beam (with
  dark blue boundaries) flows at a velocity $v_j$ parallel to the
  $z$-axis. The streaming environment (initially filling the
  region above the cyan line) moves at a velocity
  $v_a$ on the $xz$-plane, at an angle $\phi$ with respect to the
  $z$-axis. The streaming environment produces an initial
  deflection $\delta$ of the jet with respect to its initial
  propagation. The jet/environment interaction region is shown by the
  thick, red curve.
\label{fig3}}
\end{figure}

\section{The computed models}

We have used the ``yguaz\'u'' code (Raga et al. 2000) to compute
a number of ``jet/shear layer interaction'' models. In all simulations,
the 3D Euler equations are integrated in a 5-level binary adaptive
grid, covering a $(x,y,z)$ cartesian domain of a
$(x_{max},y_{max},z_{max})=(1,1,2)\times 10^{18}$~cm
physical size. This domain corresponds to $256\times 256 \times 512$
cells (with a $\sim 3.9\times 10^{15}$~cm size) at the highest grid
resolution.

The cylindrical jet beam is injected (at all times) on the $z=0$ boundary,
within a circular region of radius $r_j=5\times 10^{16}$~cm centered
on the point $(x_j,y_j)=(2,5)\times 10^{17}$~cm. The jet velocity
$v_j$ is initially parallel to the $z$-axis, with the time-dependence
described in section 3.

On the $x=0$ grid boundary, for $z>z_s=2\times 10^{17}$~cm
at all times we impose an inflow condition, with
a moving environment (of density
$n_a$, temperature $T_a$ and velocity $v_a$, directed at an angle
$\phi$ with respect to the $z$-axis) for $z>z_s$. 
Outflow conditions are imposed on the remaining grid boundaries.

The computational domain is initially filled with the moving
environment (of density $n_a$, temperature $T_a$ and velocity $v_a$,
see above) above a plane interface starting at $(x,z)=(0,z_s)$
and inclined at an angle $\phi$ with respect to the $z$-axis,
so that it is parallel to the direction of $v_a$ (see above). Below
this inclined interface, the computational domain is filled with
a stationary, uniform environment of density $n_b$ and temperature
$T_b$.

A rate/transport equation for ionized H is integrated together with the
gasdynamic equations. With the H ionization fraction, the density
and the temperature, we calculate the HI, H~II, OI and O~II cooling
rates (assuming that the ionization of O follows the H ionization),
and switch to a parametrized cooling rate for high temperatures,
as described by Raga et al. (2002). This cooling rate is included
in the energy equation.

In the initial and inflow conditions, H is neutral. A small
electron density (assumed to come from singly ionized C) is included
in order to feed the collisional ionization cascade induced by
the passage of shocks. At the $\sim 3.9\times 10^{15}$~cm resolution
of the simulations, in some of the shocks that occur in the flow
the cooling regions are unresolved by $\sim$ one order of magnitude.

\begin{deluxetable}{ccccccr}
\tablenum{1}
\tablecaption{Computed models}
\tablewidth{0pt}
\tablehead{
  \colhead{Model} & \colhead{$v_a$ [km s$^{-1}$]} &
  \colhead{$n_a$ [cm$^3$]} &\colhead{$T_a$ [K]} &
  \colhead{$n_b$ [cm$^3$]} &\colhead{$T_b$ [K]} &\colhead{$\phi$ [$^\circ$]}
}
\startdata
M1 & 30 & $2\times 10^3$ & 100 & 100 & $2\times 10^3$ & 70 \\
M2 & 30 & $2\times 10^3$ & 100 & 100 & $2\times 10^3$ & 90 \\
M3 & 30 & $2\times 10^3$ & 100 & 100 & $2\times 10^3$ & 110 \\
M4 & 10 & $2\times 10^4$ & 100 & 1000 & $2\times 10^3$ & 90 \\
M5 & 10 & $5\times 10^3$ & 100 & 1000 & $2\times 10^3$ & 110 \\
M6 & 10 & $10^4$ &         100 & 200 & $5\times 10^3$ & 110 \\
M7 & 10 & $2\times 10^4$ & 100 & 1000 & $2\times 10^3$ & 110 \\
M8 & 15 & $10^4$ &         100 & 200 & $5\times 10^3$ & 110 \\
\enddata
\end{deluxetable}

For the same inflowing jet (\S 3), we have computed 8 simulations
with the environmental characteristics shown in Table 1:
\begin{itemize}
\item models M1-M3 have the same densities, temperatures
and velocities for the regions above and below the environmental
shear layer (see Table 1), and illustrate the effect of
changing the orientation of the environmental shear layer with respect to
the jet. These three models have a streaming environment with
an environment to initial jet density ratio $n_a/n_j=20$,
mass flux ratio $n_av_a/n_jv_j=2.4$
and a ram-pressure ratio $p^2=n_av_a^2/n_jv_j^2=0.29$.
\item model M4 has
the same $\phi=90^\circ$ shear layer orientation as model
M2, but has a higher environment to jet density
ratio $n_a/n_j=200$, a mass flux ratio $n_av_a/n_jv_j=8.0$,
and a $n_av_a^2/n_jv_j^2=0.32$ ram pressure
ratio (similar to the one of models M1-M3),
\item models M5-M7 all share a $\phi=110^\circ$ orientation angle
  between the jet and the environmental shear layer, and a
  $v_a=10$~km~s$^{-1}$ velocity shear. The models are an exploration
  of the effect of varying the shearing environment density $n_a$,
  with the three values given in Table 1, resulting in
  $n_a/n_j=50$, 100 and 200 density ratios,
  $n_av_a/n_jv_j=2.0$, 4.0 and 8.0 mass flux ratios, and
  $n_av_a^2/n_jv_j^2=0.08$, 0.16 and 0.32 ram pressure ratios,
  (for models M5, M6 and M7, respectively),
\item model M8 is the same as model M6, but with a $v_a=15$~km~s$^{-1}$
  velocity shear. The ram pressure of the streaming environment
  is similar to the one of model M7.
\end{itemize}

\section{Results}

\subsection{Model M2}

Of our simulations (see \S 4), model M2 has a deflection angle
similar to the HH~270/110 system, and shows small scale
structures resembling the ones observed in HH~110. We therefore choose
this model for a discussion of the time-evolution of the flow.
In this model, the variable jet (injected along the $z$-axis)
interacts with a streaming environment flowing parallel to
the $x$-axis (with a velocity $v_a=30$~km~s$^{-1}$ at a $\phi=90^\circ$
angle with respect to the impinging jet, see Table 1).

The first and second columns of Figure 4 show
the density and temperature distributions on
the $y=y_j$ plane (a cut through the central plane of the jet/streaming
environment interaction flow) at three different integration times.
The third column of Figure 4 shows column density maps, obtained
through integrations along the $y$-axis (i.e., assuming that the
$xz$-plane is parallel to the plane of the sky). The column
densities have been calculated using only the material disturbed
by the jet or by the environmental shear surface.

\begin{figure}
\epsscale{0.7}
\plotone{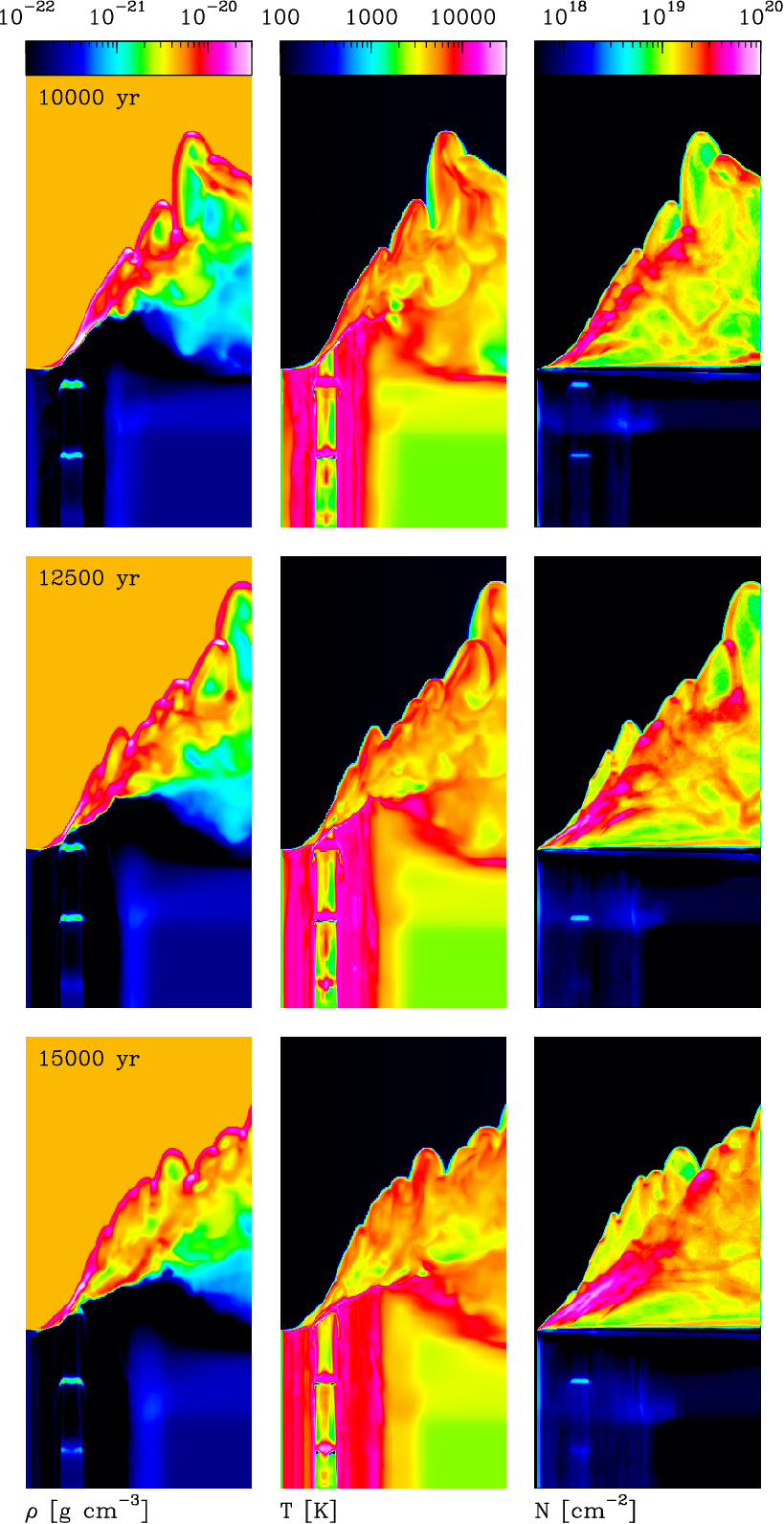}
\caption{(Left) Density (in cm$^{-3}$) and (Center) temperature (in K)
  $xz$ mid-plane cuts,
  and (Right) column density maps (in cm$^{-2}$) integrated along lines
  of sight perpendicular to the $xz$-plane (see Figure 3) for
  model M2 (see Table 1) at integration times of $10000$, $12500$
  and $15000$~yr. The regions with unperturbed environment have not
  been included in the calculation of the column densities.
  The environment is divided into an initially
  stationary, lower region (in black/blue in the temperature cuts)
  and an upper region in which the environment streams to the right
  (perpendicular to the jet direction, in the $\phi=90^\circ$ M2 model,
  see Figure 3). The internal working surfaces formed as a result
  of the ejection variability (seen as horizontal bars along the
  incident jet) interact with the streaming environment producing
  a complex structure of deflected working surfaces and wakes.
  The frames have a horizontal extent of $10^{18}$~cm, approximately
  1/2 of the physical extent of the HH~270/110 images of Figure 2.
\label{fig4}}
\end{figure}

At $t=0$, the jet starts to enter the computational grid along the
$z$-axis, and the leading head of the jet reaches the environmental shear
layer (located at $z_s=2\times 10^{17}$~cm, see \S 4) at $t\approx 1700$~yr.
The jet has a sequence of ``internal working surfaces'' produced
by its periodic ejection velocity variability (see \S 3), which
successively impact the environmental shear layer.

The time-frames shown in Figure 4 correspond to later evolutionary
times, in which a number of internal working surfaces of the jet
have already reached the shear layer, and have been deflected
by the lateral environmental flow. The $t=10000$, 12500 and 15000~yr
time frames show that the environmental shear layer produces
a time-variable but systematic deflection of $\sim 45^\circ$
of the impinging jet.

In the density and temperature slices (first and second columns
of Figure 4), we see the impinging jet entering from the bottom
boundary of the computational grid. The variability of the
jet results in the production of internal working surfaces (seen as high
density and temperature, arc-like structures) which impinge on the
environmental velocity shear layer. This jet is surrounded by a low density,
warm ($\sim 10^4$~K) cocoon produced by the initial leading working
surface of the jet as well as by the passage of the successive internal
working surfaces. In the deflected jet region, complex density
and temperature structures are obtained. Some of these structures
can be followed in the successive time frames, as they
are advected in the general direction of the deflected flow.

It is clear that the column density maps have morphologies
that resemble the optical and IR observations of HH~270/110 (shown
in Figure 2). This qualitative agreement is clearly better than the one
obtained previously with simulations of jets colliding with stationary,
dense clouds (see Raga et al. 2002).

We have also calculated proper motions from the main structures
seen in the $t=12500$~yr frame. To do this, we combine the
column densities in this time-frame with the ones obtained
for $t=12530$~yr, and obtain the motions of the higher
density structures with a standard, wavelet convolution algorithm
(see Raga et al. 2017).

Figure 5 shows a comparison of these proper motions
with the ones derived for HH~270/110 by Kajdic et al. (2012)
from ground-based H$\alpha$ images covering $\sim 15$~yr
(they obtain similar results from [S~II] images). The similarities
between the simulated and observed proper motions are quite striking.
The overall size of the structure, 5$\arcsec$ - 10$\arcsec$, and the range of tangential velocities, 50 - 200 km/s,
match reasonably well the observations.
Models of jet/stationary cloud collisions do not result in such
a satisfying agreement of predicted and observed proper motions
(see Figures 6 and 8 of Raga et al. 2002).

\begin{figure}
\epsscale{0.9}
\plotone{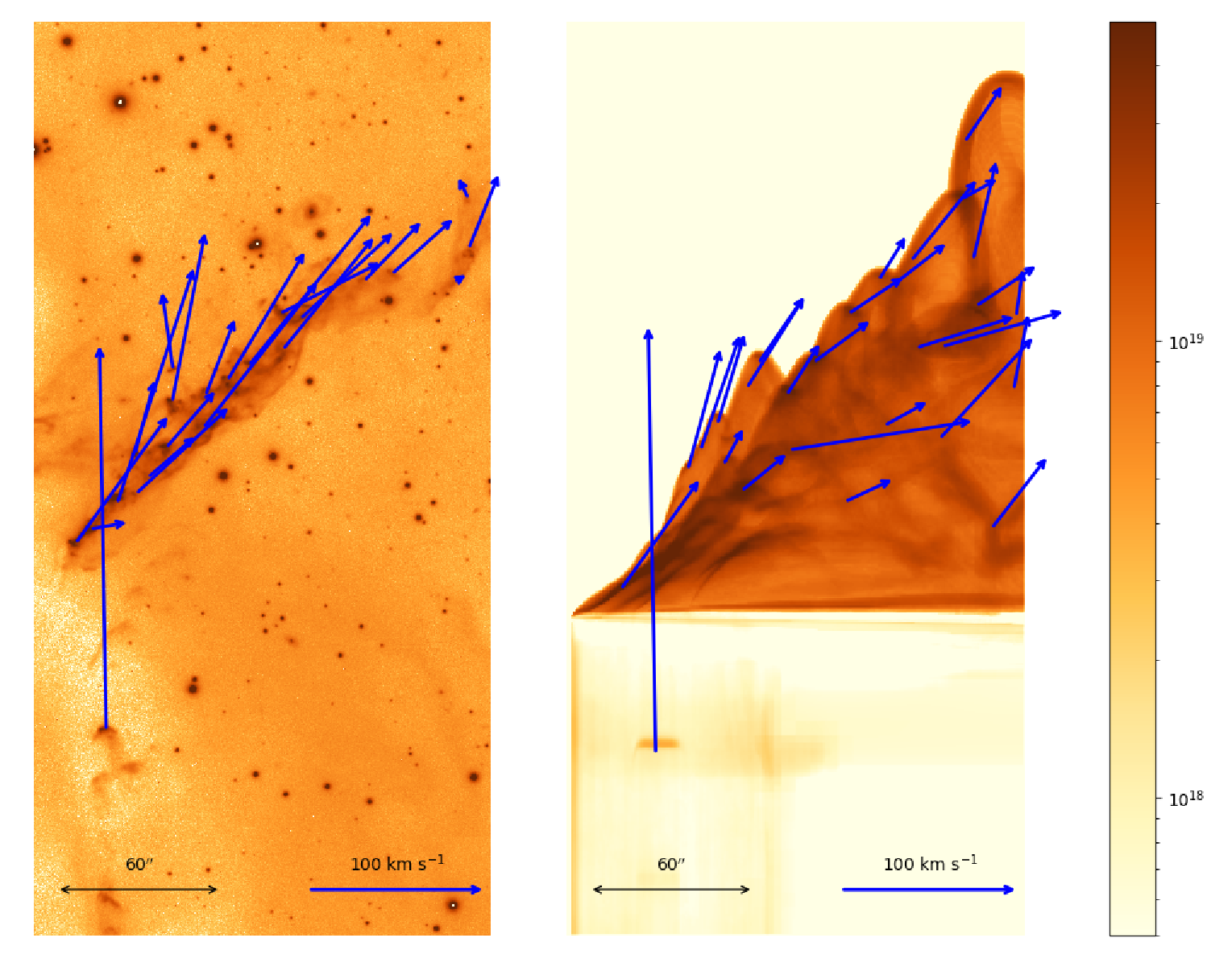}
\caption{Left: Subaru H$\alpha$+red [S~II] image of HH~270/110 (rotated
  so that the HH~270 axis is approximately parallel to the ordinate)
  and right: the $t=12500$~yr column density map obtained from model M2
  (center, right frame of Figure 3). The blue arrows show the proper
  motions measured on ground-based H$\alpha$ images by Kajdic et al. (2012),
  and the proper motions measured on two column density maps with
  a time difference of 30~yr (right). The scale of the proper motion
  velocities is given by the 100~km~s$^{-1}$ arrows on the bottom right
  of the two frames, and the angular scale (assuming a 400~pc distance)
  by the bottom left, $60''$ arrows.
\label{fig5}}
\end{figure}

\subsection{The other models}

Figures 6 thru 8 show some of the results obtained for all of the models
of Table 1:
\begin{itemize}
  \item Figures 6-7 show column density maps obtained
by integrating along lines of sight perpendicular to the mid-plane
of the oblique jet/streaming environment interaction (see Figure 2),
at integration times chosen so that the head of the deflected jet
is close to the boundary of the computational grid,
\item Figures 8 shows maps of column densities integrated
  along lines of sight parallel to the $x$-axis, displaying
  the structure of the flow perpendicular to the plane shown
  in Figure 3.
\end{itemize}
The column densities are calculated only with the jet material
and the disturbed environmental gas.

\begin{figure}
\epsscale{1.1}
\plotone{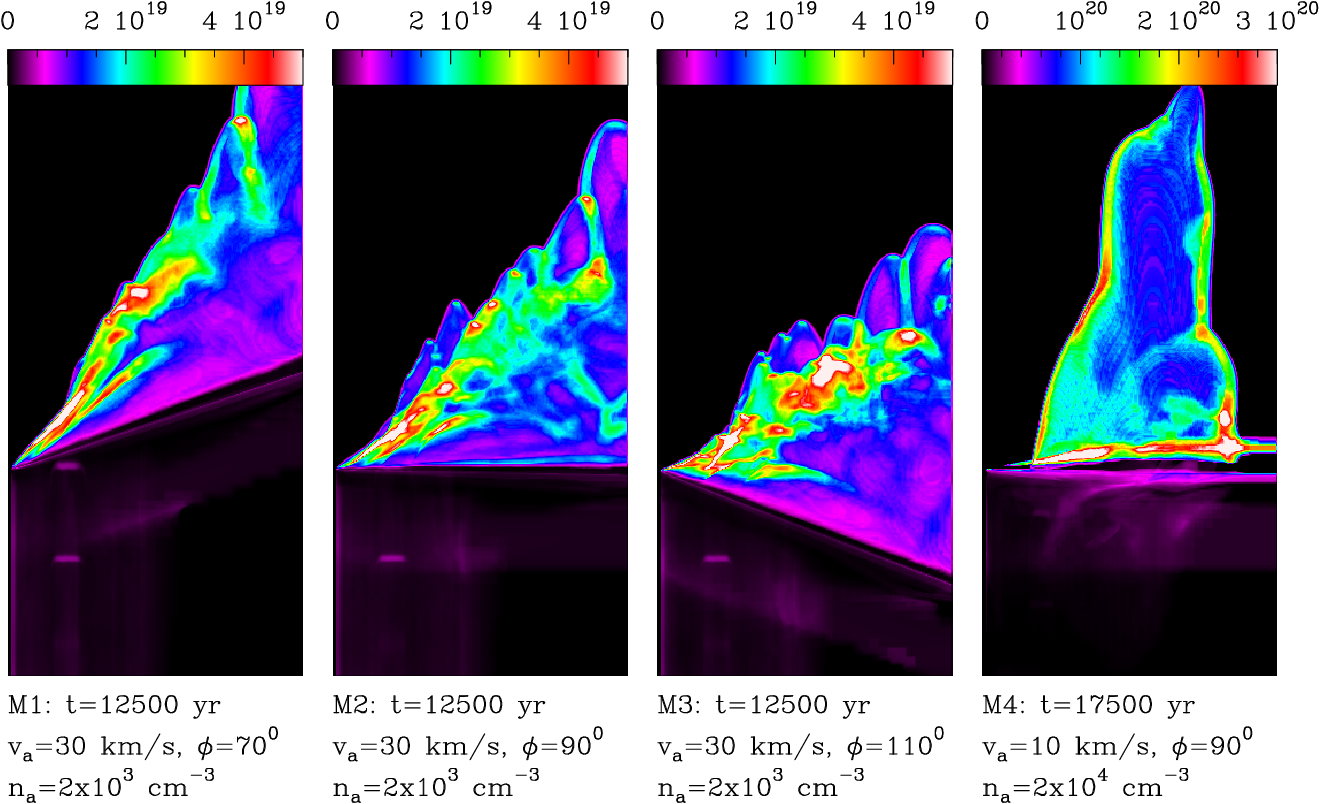}
\caption{Column density maps obtained for models M1-M4 (see Table~1)
  at times when the leading head has just left or is leaving the
  computational domain. The column densities have been calculated
  by integrating the density along the $y$-axis (perpendicular
  to the plane of Figure 3), and the unperturbed regions
  of the environment have not been included in the calculation
  of the column densities. The linear colour scale is given (in cm$^{-2}$)
  by the top bars. The whole $xz$ computational domain, of
  $(1,2)\times 10^{18}$~cm is shown. Some of the parameters of the
  models are listed below each frame.
\label{fig6}}
\end{figure}

\begin{figure}
\epsscale{1.1}
\plotone{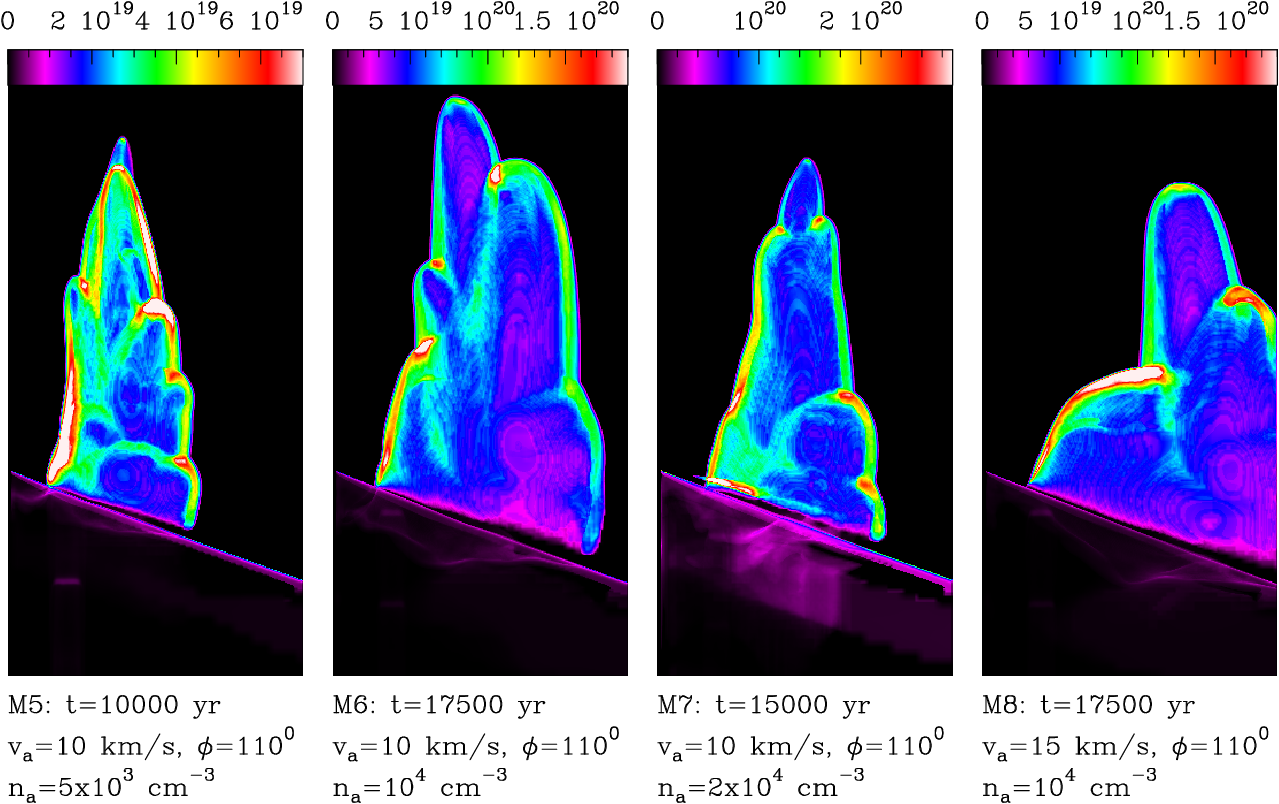}
\caption{The same as Figure 6, but for models M5-M8 (see Table 1).
\label{fig7}}
\end{figure}

While all of our models produce substantial deflections
of the jet (a result of choosing a jet-to-streaming
environment ram pressure $\sim 1$, see \S 3), we find two
distinct regimes:
\begin{itemize}
\item our ``low $n_a$'' models (models M1-M3, with $n_a=2000$~cm$^{-3}$
  and $v_a=30$~km~s$^{-1}$, see Table 1) produce highly
  structured jet/environment interaction regions, in which the individual
  working surfaces of the impinging jet penetrate the streaming
  environment, and are then advected sideways as a result of
  the interaction (see the first three frames of Figure 6),
\item the remaining models (models M4-M8, with $n_a=5\times 10^3\to
  2\times 10^4$~cm$^{-3}$ and $v_a=10\to 15$~km~s$^{-1}$, see Table 1)
  produce jet/environment interaction regions with larger scale
  structures (see the fourth panel of Figure 6, and the four
  panels of Figure 7), but lack the smaller scale ``turbulent appearance''
  of models M1-M3.
\end{itemize}

The lack of small-scale structure of models M4-M8 appears to be
the result of the higher inertia of the shocked environment, which
does not allow substantial penetration by the individual working
surfaces of the impinging jet.

\begin{figure}
\epsscale{1.0 }
\plotone{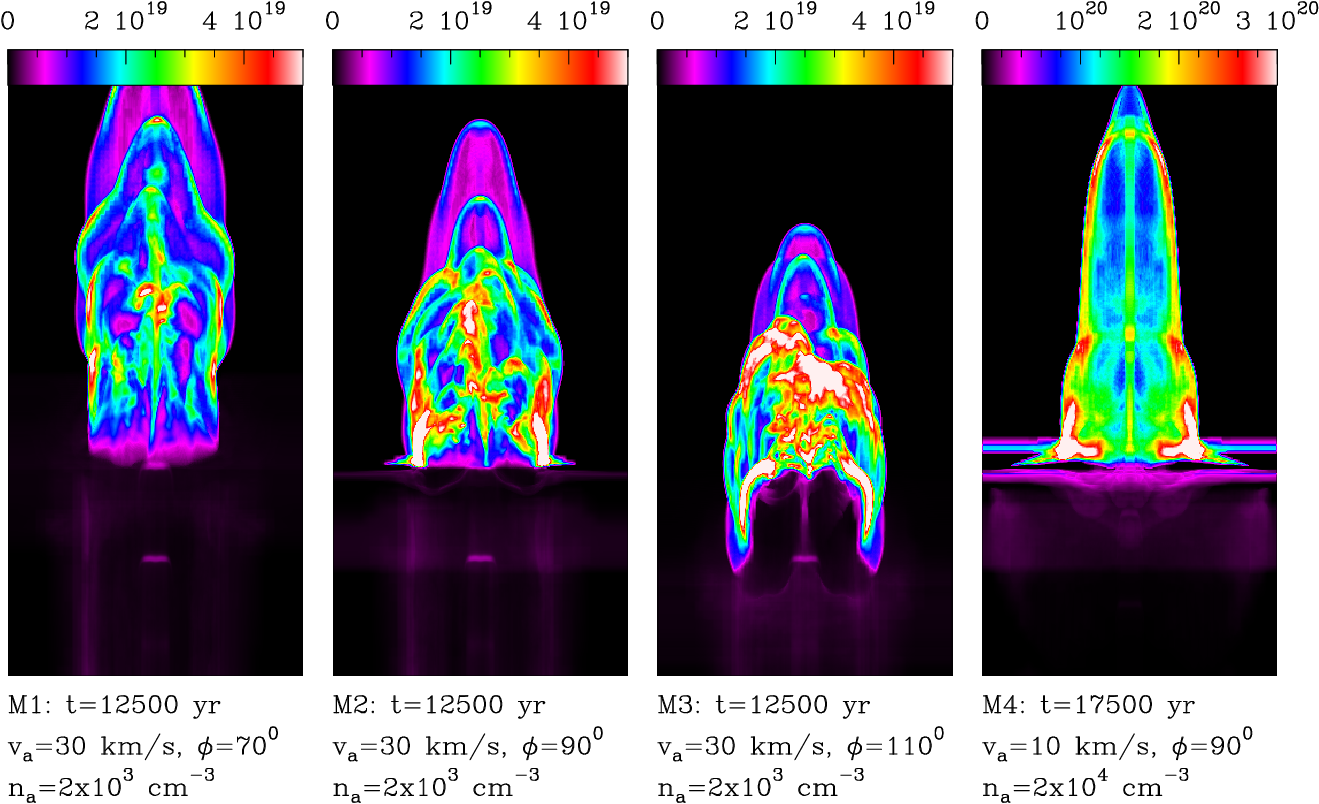}
\vskip 0.2in
\plotone{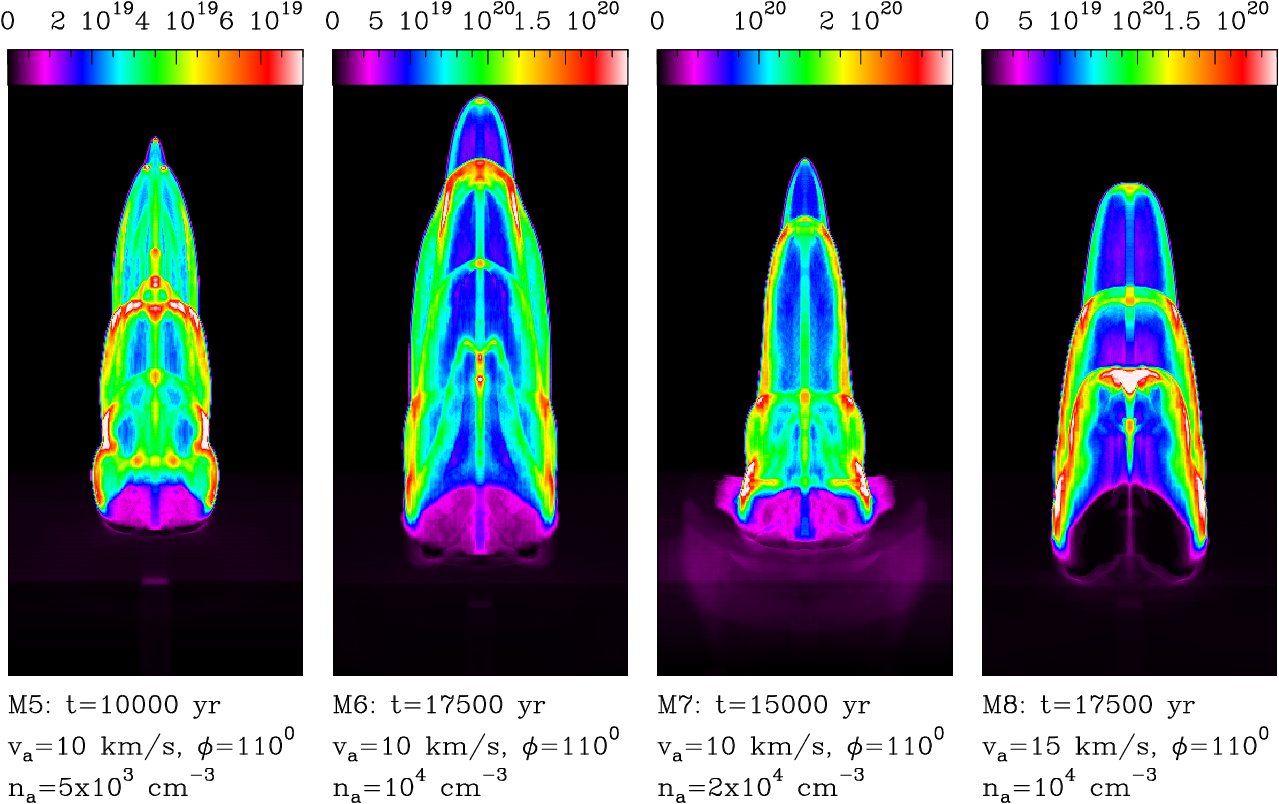}
\caption{Column density maps obtained for models M1-M4 (top) and M5-M8 (bottom) (see Table~1)
  at the same times as the frames of Figure 5. These
  maps correspond to column densities obtained
  by integrating the density along the $x$-axis, assuming that
  the plane of the sky is parallel to the $zy$-planes (i.e.,
  that it is perpendicular to the plane of Figure 3).
\label{fig8}}
\end{figure}

Even though the column density maps of Figures 6-7 correspond
to the approximate orientation of the HH~270/110 flow, we have
also computed column density maps obtained by integrating along the $x$-axis
(Figure 8). These maps do not correspond to the observed
HH~270/110 outflow, but we have included them to illustrate what
might be seen in other jet deflection systems observed with different
orientations with respect to the plane of the sky.

As expected, these column density maps (Figure 8) also show a remarkable
difference between the low $n_a$ (M1-M3, first three panels) and high
$n_a$ models (M4-M8, last five models). The low $n_a$ models show: (1) small scale structures of turbulent appearance,
(2) larger scale bow shaped ``heads'', and (3) a central, twisted beam structure.

The side-to-side asymmetries in the flow (first three panels
of Figure 8) are triggered by a one-grid point asymmetry in the initial
jet cross section.

The high $n_a$ models (models M4-M8 in Figure 8)
and the four panels at the bottom) do not show side-to-side
asymmetries (though they share the small asymmetry in the inflow
condition with models M1-M3). They do show large-scale bow like
structures.

\section{Discussion}

The HH~270/110 system is the best studied example of a possible
jet deflection through a collision with a dense obstacle. The
sudden bending of the flow, and the decrease in proper motion
velocities going from HH~270 to HH~110 are consistent with a collision
scenario (Reipurth et al. 1996).

A new JWST image of this outflow shows quite dramatically that
the working surfaces traveling along the HH~270 jet are deflected
sideways to form the HH~110 flow (see \S 2 and Figures 1 and 2).
This observation
is consistent with the scenario of Raga et al. (2023) of a steady
jet interacting with an environment with a sideways velocity shear,
but instead with a variable jet with successive internal working surfaces.

We have then run 8 numerical simulations of a variable jet hitting
an environmental velocity shear layer. In order to produce
a deflection of $\sim 45^\circ$ of the incident jet (as observed
in the HH~270/110 system), all of our models
have an environment-to-jet ram pressure ratio of $\sim 1$. All of
our simulations have the same incident jet, with parameters appropriate
for HH~270.

We first show three models with a streaming environment region of
velocity $v_a=30$~km~s$^{-1}$ and density $n_a=2\times 10^3$~cm$^{-3}$,
exploring different values of the angle $\phi$ between the
jet and the streaming environment velocities (see Figure 3). These
three models show column density
structures that qualitatively resemble the HH~270/110
system, with incident working surfaces that are deflected by the
interaction with the streaming environment, forming an outer
envelope of the interaction region. Also,
in the wake of these working surfaces there is a broader region filled
with a complex structure of peaks in the computed column densities
(see Figures 4 and 6). Such a region is also seen in optical images
of HH~110 (see Figure 2).

Even though the simulations with angles $\phi=70$, 90 and $110^\circ$
between the jet and the shear layer (models M1-M3, see Table 1)
all resemble HH~110, we have chosen the $\phi=90^\circ$ (model M2, see
table 1) to compute proper motion velocities of the observed column
density peaks. These velocities very strongly resemble the proper motions
of the HH~270/110 flow (see Figure 5).

If the flow parameters of models M1-M3 are indeed appropriate
for the HH~270/110 system,
they imply that the HH~270 jet is encountering a region with a quite
high velocity of $\sim 30$~km~s$^{-1}$. Such a velocity is too
high for a turbulent molecular cloud (with typical velocities
of a few km~s$^{-1}$) but might be
appropriate for a molecular environment that is strongly stirred
up by other outflows. This might be the case for HH~270/110, which
has two outflows to the N and W of HH~110 (see Figures 8 and 10
of Kajdic et al. 2012).

We have then computed five other simulations with lower velocities
for the shearing environment ($v_a=10$ and 15~km~s$^{-1}$, models
M4-M8, see Table 1 and Figures 6-7). In order to have an environment
with a ram-pressure similar to the ram-pressure of the jet (necessary
for producing a substantial jet deflection), the environmental
density then has to be substantially higher (in the $n_a=5\times 10^3\to
2\times 10^4$~cm$^{-3}$ range). These lower $v_a$, higher $n_a$ models produce
column density maps that do not have the small-scale structures seen
in the deflected region of the HH~270/110 outflow.

From this limited exploration of the parameter space appropriate
for modeling HH~270/110, we tentatively conclude that the deflection
observed in this system is due to an interaction with a relatively
high velocity ($\sim 30$~km~s$^{-1}$), streaming environment. Such
a velocity could possibly be associated with perturbations from
other outflows in the region, and in particular from HH~451. This object
might be part of a large outflow composed by HH~491, 451 and 490, from
N to S, see Figure 8 of Kajdic et al. (2012). These authors have
already suggested that HH~451 might be responsible for the HH~270/110
deflection. It is also interesting to notice that laboratory experiments using high power lasers (Yuan et al. 2015)
do show an interacting morphology and deflection angle consistent with this explanation.
One needs to be cautious to over interpret the models, since the models do not encompass the full flow history, including the large scale precession that is partially observed in wide file images like those from Spitzer or WISE.

\section{Appendix}

\subsection{Wide Field View of the HH~270/HH~100 Systems}
 The HH~270/HH~110 is a remarkable system and there is a wide range of observations, both from the ground and space that can shed some light in the interpretation of the flow history.
 Figure 9 (Left), shows the Spitzer 8 $\mu$m image, similar to Fig 1, except that the extinction regions are the clearest, in particular around HH~270 and IRAS 0587+0255, the driving source of HH~451/HH~490/HH~491 outflow. We have drawn 4 arcs to show that the flow is not fully symmetric between the NE and SW, as one naively expects from precession, and that a 'break' of symmetry takes place around HH~110, and that arc follows a closer path towards HH~490. Finally the red arc marks the location of HH~1034, that Kajdic et al. (2012) suggested to be in along the same overall direction as HH~102 in the NE. If a large precessing flow gave birth to both condensations, it suggests that the formation of HH~110 is perhaps a more recent event. Star formation does not take place in isolation, one only needs to look at the complexity of multiple systems in Lynds 1641 (e.g. Morgan et al. 1991; Galfalk \& Olofsson 2007), NGC 1333 (Plunkett et al. 2015; Langeveld et al. 2024), HH~24 (Reipurth et al. 2023) or Serpens (Green et al. 2024), to realize that interaction among nearby proto-stellar outflows is quite possible.

{\begin{figure}[b!]
\epsscale{1.1}
\plotone{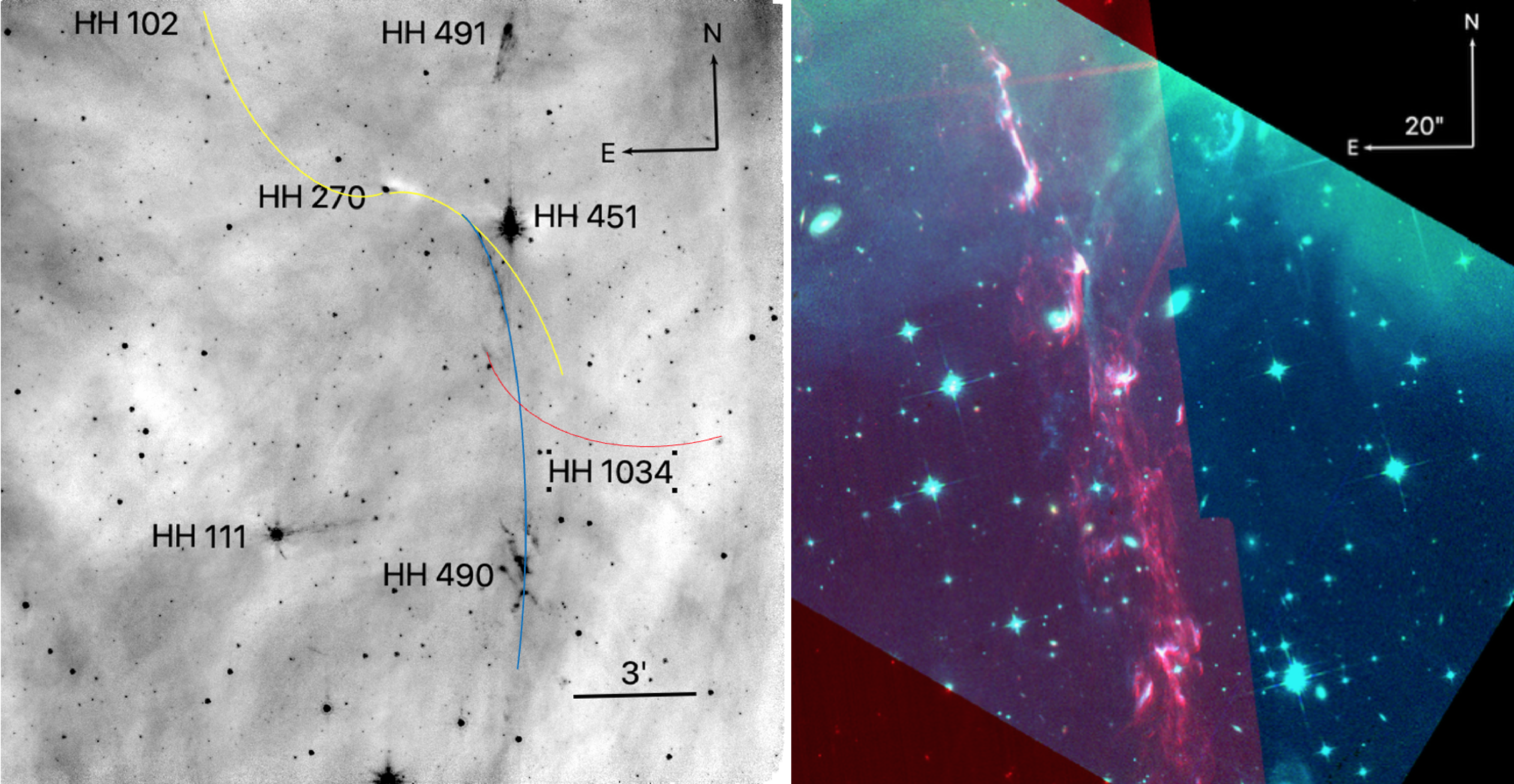}
\caption{Spitzer IRAC Channel 4 map of HH~110/HH~111 at 8$\mu$m (left), emphasizing the dark higher extinction and gas column density regions around the HH~270 and IRAS 05487+0255 sources. The yellow arcs describe the suggested precession path that reaches HH~112 in the NE and HH~490 in the SW. (Right) A three color image using WFC3 F110W (blue), F160W (green) and NIRCam F460M. It shows the intertwine interaction between the emission arising from HH~451 and HH~110.
\label{fig10}}
\end{figure}

\subsection{HST and JWST data Comparison}

 A et of observations at the high angular resolution, comparable to JWST, comes from the Wide Field Camera 3 (WFC3) on board of the Hubble Space Telescope (HST). Figure 9 (right) shows a composite three color image with data obtained with the F110W (blue) and  F160W (green) filters and the NIRCam F460M (red) filter. The F110W bandpass [0.9 – 1.4$\mu$m] is sensitive to some of the Hydrogen Paschen lines (Pa~$\beta$ 1.28$\mu$m, Pa~$\gamma$ 1.09$\mu$m and Pa~$\delta$ 1.00$\mu$m). The F160W bandpass [1.4-1.7$\mu$m] could contain a fraction of the well known shock excited [Fe~II] 1.64$\mu$m line. One of the remarkable features of the comparison is The arc arising from HH~451 source, presumably tracing the slow expanding cavity of its outflow is clearly engulfed by the H2 flow.
 
\subsection{HH~270 and HH~110 knots} 

As mentioned in the introduction, the HH~270/HH~110 system has been studied in great detail by Kajdic et al. (2012). In their manuscript, that relied on ground based observations,they identified several knots and/or condensations plus their proper motions were measured. For context in this section we have included a closer look (Fig 10) at the NIRCam F460M image and have tried to match some of the knots identified by Kajdic et al. (2012) following their nomenclature (see their Fig 5 and Table 3). The superb angular resolution of JWST shows a more complex structure and makes a bit more complicated a one-to-one comparison.

\begin{figure}
\epsscale{1.2}
\plotone{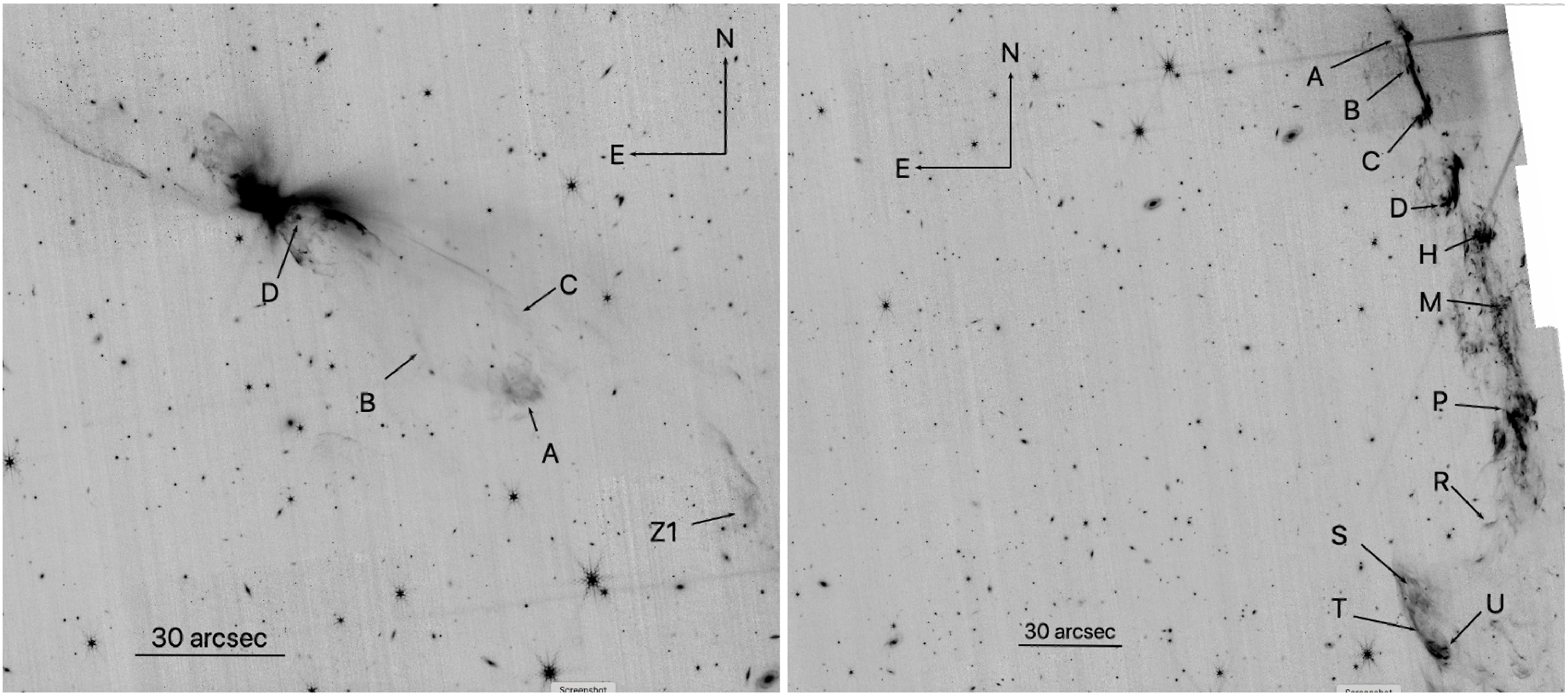}
\caption{A closer look at the NIRCam F460M (Fig 1). On the left the HH~270 outflow with some of the knots identified by Kajdic et al. (2012). On the right the HH~110.
\label{fig10}}
\end{figure}
\vskip 0.5in

This work is based in part on observations made with the NASA/ESA/CSA James Webb Space Telescope. The data were obtained from the Mikulski Archive for Space Telescopes (MAST) at the Space Telescope Science Institute, which is operated by the Association of Universities for Research in Astronomy, Inc., under NASA contract NAS 5-03127 for JWST. The specific observations analyzed can be accessed via  DOI \dataset[10.17909/69x5-m651]{https://doi.org/DOI}
The mentioned WFC3 observations can be accessed via DOI \dataset[10.17909/8r4m-xx27]( http://dx.doi.org/10.17909/8r4m-xx27)

\begin{acknowledgments}
 Dr. Alejandro Raga, lead author of this work, passed away on July 20, 2023.
A couple of weeks before the co-authors provided the final comments to update the manuscript prior submission. Dr. Raga was an amazing friend and colleague, he is fondly remembered and sadly missed.
\vskip 0.05in   
\end{acknowledgments}
\vskip 0.05in

$*$ {\bf Contact Author}


\begin{acknowledgments}
  MEL \& CAP  acknowledge support from NASA EPSCoR grant 80NSSC22M0167.
  AR and JC acknowledge support from the DGAPA (UNAM) grant IG100422.
  A C-R acknowledges support from a CONACYT postdoctoral fellowship.
  This research was carried out in part at the Jet Propulsion Laboratory,
  which is operated by the California Institute of Technology under a
  contract with the National Aeronautics and Space Administration
  (80NM0018D0004).
\end{acknowledgments}




\vskip 1in

\begin{thebibliography}

\bibitem[de Gouveia Dal Pino(1999)]{1999ApJ...526..862D} de Gouveia Dal Pino, E.~M.\ 1999, \apj, 526, 862. doi:10.1086/308037

\bibitem[Galfalk \& Olofsson (2007)]{2007A&A..466...579} Galfalk, M. \& Olofsson, G.\ 2007, \aap, 466, 579. doi:10.1051/0004-6361:20066852

\bibitem[Gardner et al. (2023)]{2023PASP..135f8001G} Gardner, J.P., Mather, J.C., Abbott, R. et al. 2023, \pasp, 135f8001G. doi: 10.1088/1538-3873/acd1b5 

\bibitem[Green, J. et al. (2024)]{2024, JWST Press release  ID: 2024-115} Green et al. 2024, JWST Press Release ID: 2024-115 (https://webbtelescope.org/contents/news-releases/2024/news-2024-115)

\bibitem[Kajdi{\v{c}} et al.(2012)]{2012AJ....143..106K} Kajdi{\v{c}}, P., Reipurth, B., Raga, A.~C., et al.\ 2012, \aj, 143, 106. doi:10.1088/0004-6256/143/5/106

\bibitem[Langeveld et al. (2024)]{2024AJ..168..179} Langeveld A. B. et al. \ 2024, \aj, 168, 179. doi:10.3847/1538-3881/ad6f0c 

\bibitem[L{\'o}pez et al.(2005)]{2005A&A...432..567L} L{\'o}pez, R., Estalella, R., Raga, A.~C., et al.\ 2005, \aap, 432, 567. doi:10.1051/0004-6361:20041402

\bibitem[Morgan et al. (1991)]{1991ApJ...376..618} Morgan, J,A. et al. \ 1991, \apj, 376, 618. doi:10.1086/170310

\bibitem[Nony et al.(2020)]{2020A&A...636A..38N} Nony, T., Motte, F., Louvet, F., et al.\ 2020, \aap, 636, A38. doi:10.1051/0004-6361/201937046

\bibitem[Noriega-Crespo et al.(1996)]{1996ApJ...462..804N} Noriega-Crespo, A., Garnavich, P.~M., Raga, A.~C., et al.\ 1996, \apj, 462, 804. doi:10.1086/177195

\bibitem[Noriega-Crespo(2017)]{2017jwst.prop.1293N} Noriega-Crespo, A.\ 2017, JWST Proposal. Cycle 1, 1293

\bibitem[Plunkett et al. (2015)]{2015ApJ...803..22} Plunkett et al. \ 2015, \apj, 803, 22. doi:10.1088/0004-637X/803/1/22 

\bibitem[Raga \& Biro (1993)]{1993MNRAS...264..758} Raga A.~C. \& Biro, S. \ 1993, \mnras, 264, 758. doi:10.1093/mnras/264.3.758

\bibitem[Raga et al.(1990)]{1990ApJ...364..601R} Raga, A.~C., Cant{\'o}, J., Binette, L., et al.\ 1990, \apj, 364, 601. doi:10.1086/169443

\bibitem[Raga, Canto \& Biro (1993)]{1993MNRAS..260..163} Raga, A.~C., Cant{\'o}, J. \& Biro, S. \ 1993, \mnras, 260 163. doi:10.1093/mnras/260.1.163

\bibitem[Raga \& Canto(1995)]{1995RMxAA..31...51R} Raga, A.~C. \& Cant{\'o}, J.\ 1995, \rmxaa, 31, 51 
  
\bibitem[Raga \& Cant{\'o}(1998)]{1998RMxAA..34...73R} Raga, A.~C. \& Cant{\'o}, J.\ 1998, \rmxaa, 34, 73

\bibitem[Raga et al.(2002)]{2002A&A...392..267R} Raga, A.~C., de Gouveia Dal Pino, E.~M., Noriega-Crespo, A., et al.\ 2002, \aap, 392, 267. doi:10.1051/0004-6361:20020851

\bibitem[Raga et al.(2013)]{2013AJ....145...28R} Raga, A.~C., Noriega-Crespo, A., Carey, S.~J., et al.\ 2013, \aj, 145, 28. doi:10.1088/0004-6256/145/2/28

\bibitem[Raga et al.(2017)]{2017RMxAA..53..485R} Raga, A.~C., Reipurth, B., Esquivel, A., et al.\ 2017, \rmxaa, 53, 485. doi:10.48550/arXiv.1708.01585

\bibitem[Raga et al.(2023)]{2023RMxAA..59..105R} Raga, A.~C., Castellanos-Ram{\'\i}rez, A., \& Cant{\'o}, J.\ 2023, \rmxaa, 59, 105. doi:10.22201/ia.01851101p.2023.59.01.07

\bibitem[Ray et al. (2023)]{2023Nature...622..48R} Ray, T. P.; McCaughrean, M. J.; Caratti o Garatti, A.; Kavanagh, et al. 2023, Nature, 622, 48R. doi:10.1038/s41586-023-06551-1

\bibitem[Reipurth \& Olberg(1991)]{1991A&A...246..535R} Reipurth, B. \& Olberg, M.\ 1991, \aap, 246, 535

\bibitem[Reipurth et al.(1996)]{1996A&A...311..989R} Reipurth, B., Raga, A.~C., \& Heathcote, S.\ 1996, \aap, 311, 989

\bibitem[Reipurth et. al. (2023)] {2023ApJ..165...209} Reipurth, B., Bally, J., Yen, H-W. et al. \ 2023, \apj, 165, 209. doi: 
10.1088/1538-3873/acd1b5 

\bibitem[Rieke et al.(2023)]{2023PASP..135b8001R} Rieke, M.~J., Kelly, D.~M., Misselt, K., et al.\ 2023, \pasp, 135, 028001. doi:10.1088/1538-3873/acac53

\bibitem[Rigby et al.(2023)]{2023PASP..135d8001R} Rigby, J., Perrin, M., McElwain, M., et al.\ 2023, \pasp, 135, 048001. doi:10.1088/1538-3873/acb293

\bibitem[Riera et al.(2003a)]{2003A&A...400..213R} Riera, A., L{\'o}pez, R., Raga, A.~C., et al.\ 2003, \aap, 400, 213. doi:10.1051/0004-6361:20021879

\bibitem[Riera et al.(2003b)]{2003AJ....126..327R} Riera, A., Raga, A.~C., Reipurth, B., et al.\ 2003, \aj, 126, 327. doi:10.1086/375759

\bibitem[Wright et al.(2023)]{2023PASP..135d8003W} Wright, G.~S., Rieke, G.~H., Glasse, A., et al.\ 2023, \pasp, 135, 048003. doi:10.1088/1538-3873/acbe66

\bibitem[Yuan, D. et al. (2015)]{2015ApJ...815,46Y} Yuan, Dawei; Wu, Junfeng; Li, Yutong et al. \ 2015, \apj, 815, 46. doi: 10.1088/0004-673X/815/1/46

\end{thebibliography}
\end{document}